# 3D Numerical investigation of a rounded corner square cylinder for supercritical flows

Nivedan Vishwanath[1a], Aditya K. Saravanakumar[1b], Kush Dwivedi[1c], Kalluri R.C. Murthy[1d], Pardha S. Gurugubelli[1e] and Sabareesh G. Rajasekharan[*1,2]

*Department of Mechanical Engineering, BITS Pilani Hyderabad Campus, Hyderabad, India 500078*

**Abstract.** Tall buildings are often subjected to steady and unsteady forces due to external wind flows. Measurement and mitigation of these forces becomes critical to structural design in engineering applications. Over the last few decades, many approaches such as modification of the external geometry of structures have been investigated to mitigate wind-induced load. One such proven geometric modification involved the rounding of sharp corners. In this work, we systematically analyze the impact of rounded corner radii on the reducing the flow-induced loading on a square cylinder. We perform 3-Dimensional (3D) simulations for high Reynolds number flows (Re=1 × 105) which are more likely to be encountered in practical applications. An Improved Delayed Detached Eddy Simulation (IDDES) method capable of capturing flow accurately at large Reynolds numbers is employed in this study. The IDDES formulation uses a k-ω Shear Stress Transport (SST) model for near-wall modelling that prevents mesh-induced separation of the boundary layer. The effects of these corner modifications are analyzed in terms of the resulting variations in the mean and fluctuating components of the aerodynamic forces compared to a square cylinder with no geometric changes. Plots of the angular distribution of the mean and fluctuating coefficient of pressure along the square cylinder's surface illustrate the effects of corner modifications on the different parts of the cylinder. The windward corner's separation angle was observed to decrease with an increase in radius, resulting in a narrower and longer recirculation region. Furthermore, with an increase in radius, a reduction in the fluctuating lift, mean drag, and fluctuating drag coefficients has been observed.

**Keywords:** long-span suspension footbridge; galloping instability; web opening; Den Hartog criterion; damping

## 1. Introduction

Tall structures such as buildings and wind turbines encounter unsteady pressure forces which usually have an undesirable impact on their integrity. It is understood that these forces originate due to the instabilities in the shear layer around these structures. Previously, many different approaches have been explored to mitigate such wind loading, including the manipulation of the oncoming flow, the wake behind the structures, or both. Some of the commonly investigated approaches include corner modifications (Kawai 1998, Shiraishi *et al.* 1988), construction of wind fences around the building (Rehacek *et al.* 2017, Cornelis and Gabriels 2005), and the use of escarpments (Tsai and Shiau 2011).

Corner modifications delay the separation of boundary layers in the wake and diminish vortex shedding, thereby mitigating the lift and drag forces that the cylinder experiences. A vast range of corner modifications have been studied over the years, including chamfered and rounded corners, slotted corners (Tse *et al.* 2009, Kwok *et al.* 1988), cuts, and tapers (Kawai 1998, Young-Moon *et al.* 2008, Haque 2020, Choi and Kwon 1999). In (Choi and Kwon 2003), the authors performed wind-tunnel experiments to study how the Strouhal number for rectangular cylinders was affected by geometrical changes like corner cuts. Similarly, the effect of tapered corner modifications on cross-wind displacement was studied in (Young-Moon *et al.* 2008). It was concluded that the taper effect is observed for structures with low damping ratio and a high value of reduced velocity. With regards to chamfered corner modifications, an experimental investigation of chamfered square cylinders in the presence and absence of an oscillation frequency was carried out in (Aswathy *et al.* 2015). More recently, a rectangular building model with varying degrees of chamfer was experimentally investigated in (Li *et al.* 2020). It was observed that an increase in chamfering led to a decrease in wind loading. A series of experiments presented in (Kawai 1998) concluded that among three types of corner modifications (slots, recessions, and rounding), rounded modifications were the most effective in suppressing the aeroelastic instability for a square prism. It was also observed that the wind-induced vibration reduced when the corner radius was increased.

One of the earliest attempts to experimentally study the effects of corner modifications is described in (Delany and

---

*Corresponding author, Associate Professor
 E-mail: sabareesh@hyderabad.bits-pilani.ac.in
[a] Undergraduate Student
[b] Undergraduate Student
[c] Undergraduate Student
[d] Assistant Professor
[e] Assistant Professor

Sorensen 1953). The results from the study pointed at a sharp jump in the mean drag force of a corner modified cylinder with a normalised corner radius of 0.167 at Re=7 × $10^5$. Furthermore, a drag-crisis was observed, a phenomenon in which the time-averaged drag gradually decreases as the Re is increased. The mean drag eventually reaches a point of minima before increasing again. Before the minima, the flow lies in the subcritical regime. After crossing the minima, the flow becomes supercritical in nature. In the study by Lee (Lee 1975), the mean and fluctuating fields across a 2D square prism at Re=1.76 × $10^5$ were measured. Furthermore, experiments were performed by Okamoto and Uemura (Okamoto and Uemura 1991) to study the flow around a cube with varying corner radii at Re=4.74 × $10^4$. When the corner radius was increased, the authors observed a reduction of drag force experienced by the cube. A larger corner radius also led to increased flow reattachment, as also observed in (Tamura and Miyagi 1999). The results highlighted the contribution of inflow turbulence to flow reattachment on the upper and lower walls of the cylinder. A detailed study on the near wake behaviour for a corner modified square cylinder is presented in (Hu et al. 2006). The flow was experimentally studied for corner radii of 0, 0.157, 0.236, and 0.5 at Re=2600 & 6000 using particle imaging velocimetry (PIV). The results indicated that the strength of the shed vortices diminished when the r/D value was increased from 0 (corresponding to a square cylinder with no modifications) to 0.5 (corresponding to a simple circular cylinder). Additionally, the length of vortex formation doubled, which suggested that the leading edge took precedence over the trailing edge in influencing the structure of the near wake region at low Re (Jaiman et al. 2015). More recently, Carassale (Carassale et al. 2014) focused on the transition of turbulent flow around square cylinders from subcritical to the supercritical regime. Three cases with r/D={0.067, 0.133} were considered and experiments were conducted for (1.7 × $10^4$ ≤ Re ≤ 2.3 × $10^5$). In addition to the other observations, the effect of different angles of incidence on the aerodynamic performance of the cylinders was also analysed.

While experimental studies give information about the aerodynamic characteristics in the wake of these cylindrical structures, they cannot be used to study the complex flow interactions produced as a result of the corner modifications. Therefore, in addition to experiments, numerical studies have been conducted to better understand flow interactions with cylinders. Shi (Shi et al. 2018) simulated the flow around a geometrically modified square cylinder at Re=2600 using 2D and 3D Large Eddy Simulations (LES). While both methods predicted large-scale vortex formation with reasonable accuracy, only 3D LES was able to capture the forces and recirculation length satisfactorily. A comprehensive analysis on flow separation and the development of transitional flow around corner-modified cylinders for Re = 1000 was done in (Zhang and Samtaney 2016) by performing direct numerical simulations (DNS). The results indicated that the time-averaged length of the recirculation region was highest for a normalised corner radius of r/D = 0.125. Furthermore, A study on the impact of the incidence angle on the drag characteristics of a rounded corner square cylinder was presented in (Miran and Sohn 2015). The drag coefficient attained minimum value when the angle of incidence was between 5 and 10 degrees. Cao and Tamura (Cao and Tamura 2017) simulated subcritical and supercritical flows around square cylinder with a normalised corner radius r/D=0.167. Two Reynolds numbers, Re=2.2 × $10^4$ and 1.0 × $10^6$ were considered, denoting the subcritical and supercritical regimes, respectively. A decrease in the time-averaged drag value and an increase in the Strouhal number was noted for the supercritical case. They further extended their work by studying the shear-inflow effects for the above simulated cases in (Cao and Tamura 2018b) and (Cao and Tamura 2018a). Other numerical studies on rounded corner cylinders are summarised in (Wang and Gu 2015, Park and Yang 2016, Dai et al. 2017, Chiarini and Quadrio 2022).

To observe complex flow interactions in the cylinder wake, high fidelity numerical methods are commonly used. These include DNS and LES. They yield satisfactory results for the flow in the wake and can also resolve wall interactions with good accuracy (Ono and Tamura 2008, Rodriguez et al. 2015, Yeon et al. 2016). However, these methods require extremely fine grids and their degree of refinement increases with an increase in Re. This makes them computationally expensive and often unfeasible for high Re simulations. The detached eddy simulation (DES) is an intermediate scheme that offers a solution accuracy similar to LES at a modest computational cost. Some recent studies have used the DES method to study flow around corner modified cylinders (Chen 2018).

Owing to the high computational cost, there are very few studies that deal with the flow around bluff bodies at large Reynolds numbers. Furthermore, most of them consider very few corner modified configurations for their study. The present study aims to provide a systematic analysis of the changes produced in the aerodynamic behaviour of square cylinders due to geometric modifications. A Reynolds number of Re=1 × $10^5$ has been considered to study the problem in a practical context. Five cases with increasing normalised corner radius (also referred to as rounding radius) r/D=0, 0.05, 0.10, 0.15, and 0.20 are simulated in this study. Here, r is the rounding radius of the corners, while the cylinder diameter is denoted by D. An improved delayed detached eddy simulation (IDDES) formulation has been used to capture flow separation with high accuracy at a reasonable computational cost. The results from the sharp corner square cylinder simulation are validated against published LES results, thereby proving the applicability of DES to such problems. A strong emphasis is placed on understanding the effect of different corner radii on the force coefficients and the mean flow behaviour. The work analyses the following things: (1) The trend of force coefficients on the cylinder with varying corner radii; (2) The variation of mean and fluctuating values of the pressure coefficient along the surface of the cylinder; (3) The flow separation and recirculation region patterns as a consequence of corner modifications. In Section 2, the IDDES formulation employed in the current simulations is described and we discuss the problem

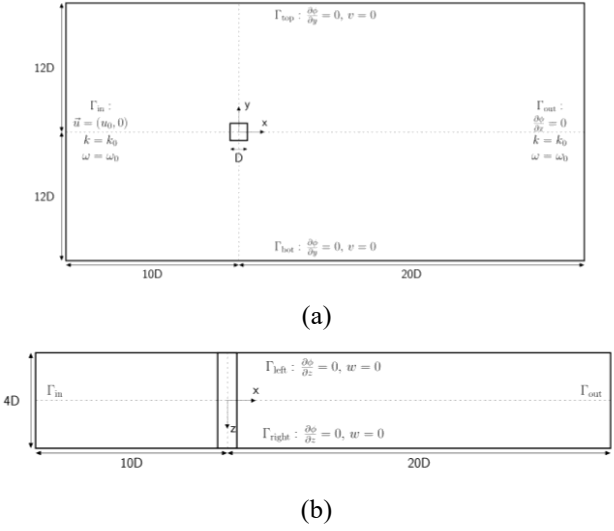

Fig. 1 2D representation of the computational domain showing the (a) side view and (b) top view. The boundary conditions imposed on each boundary $\Gamma_i$ of the domain is specified

statement in terms of the computational domain, mesh, and boundary conditions. This is followed by Section 3, which presents a discussion on the observations based on the study by comparing the resulting forces of the cylinder and flow patterns for different corner radii. Finally, Section 4 briefly summarizes the findings from this study.

## 2. Numerical method

DES is a modified approach that employs Reynolds-Averaged Navier-Stokes (RANS) models for near-wall modelling and switches to LES when the flow becomes detached. Since the instability in the flow often dominates over a small part of the computational domain, this method yields reasonably accurate results at high Reynolds numbers at a modest computational cost. However, in the standard DES model, the flow is prone to grid-induced separation (GIS) due to inadequate mesh refinement near the walls or a growing boundary layer. This is also responsible for modelled stress depletion (MSD) (Spalart *et al.* 2006) which together make up the two main drawbacks of the DES model. The delayed DES (DDES) (Menter and Kuntz 2004) tackles these issues utilizing a modified DES length scale to model the entirety of the boundary layer region using RANS models. More recently, the improved delayed DES (IDDES) was proposed as a combination of DDES and wall-modelled LES (WM-LES). It further helps with preventing GIS by increasing the modelled stress contribution across the interface. This IDDES formulation, along with the k − ω SST RANS model (Menter 1994) has been used in this work.

### 2.1 Governing equations

The RANS equations are typically employed to model incompressible, viscous fluids in the turbulent regime. These equations are given as

$$\frac{\partial \rho}{\partial t} + \frac{\partial (\rho u_i)}{\partial x_i} = 0, \quad (1)$$

$$\frac{\partial (\rho u_i)}{\partial t} + \frac{\partial (\rho u_i u_j)}{\partial x_j} = -\frac{\partial p}{\partial x_i} + \frac{\partial}{\partial x_j}\left[\mu\left(\frac{\partial u_i}{\partial x_j} + \frac{\partial u_j}{\partial x_i} - \frac{2}{3}\delta_{ij}\frac{\partial u_l}{\partial x_l}\right)\right] + \frac{\partial}{\partial x_j}\left(-\overline{\rho u_i' u_j'}\right), \quad (2)$$

where the subscripts $i$ and $j$ represent the X and Y cartesian components of the mean fluid velocity vector $\vec{u}$. The mean pressure, density, and the dynamic viscosity of the flow is given by p, ρ, and μ, respectively. In order to obtain a closed-form of the above equation, the Reynolds stress term ($-\overline{\rho u_i' u_j'}$) must be solved. The expressions for the turbulent kinetic energy k and the specific dissipation rate ω are given as

$$\frac{\partial \rho k}{\partial t} + \nabla \cdot (\rho \vec{u} k) = \nabla \cdot [(\mu + \mu_T \sigma_k)\nabla k] + P_k - \frac{\rho k^{3/2}}{l_{IDDES}} \quad (3)$$

$$\frac{\partial \rho \omega}{\partial t} + \nabla \cdot (\rho \vec{u} \omega) = \nabla \cdot [(\mu + \mu_T \sigma_\omega)\nabla \omega] + (1 - F_1)2\rho\sigma_{\omega,2}\frac{\nabla k \cdot \nabla \omega}{\omega} + \alpha\frac{\omega}{k}P_k \quad (4)$$
$$- \beta\rho\omega^2$$

$$\mu = \frac{\rho}{max(a_1 \cdot \omega, F_2 \cdot S)}, \quad (5)$$

where the turbulent viscosity is given by $\mu_T$. Gradients in mean velocity give rise to turbulent kinetic energy, the production term for which is denoted by $P_k$. The turbulent Prandtl numbers are presented by $\sigma_k$ and $\sigma_\omega$, and F1, F2 are the blending functions. Furthermore, $\sigma_{\omega,2}$, α and β are constants associated with the cross-diffusion, generation of ω and dissipation of ω, respectively. Finally, S gives the value of the strain rate tensor. The IDDES length scale is defined as functions of the RANS and LES lengths scales, $L_{RANS}$ and $L_{LES}$ as,

$$L_{IDDES} = (1 - \tilde{f}_d)L_{LES} + \tilde{f}_d(1 + f_e)L_{RANS} \quad (6)$$

$$L_{RANS} = \frac{k^{1/2}}{\beta^*\omega} \quad \text{and} \quad L_{LES} = C_{DES} \quad (7)$$

where CDES and β∗ are empirical constants and Δ is defined as Δ = min [max ($C_w$d, $C_w\Delta_{max}$, $\Delta_{min}$), $\Delta_{max}$] with d as the near-wall distance, $\Delta_{min}$ = min(Δx, Δy, Δz) and $\Delta_{max}$ = max(Δx, Δy, Δz). A detailed formulation of the functions $\tilde{f}_d$ and $f_e$ can be found in (Shur *et al.* 2008).

### 2.2 Computational model

The present study focuses on the fluid flow around a square cylinder of diameter D, modified with rounded corners of radius r. The numerical method described in Section 2 was first validated against established numerical results obtained for the flow past a square cylinder with sharp corners at Re=1 × $10^5$ using LES (Sohankar 2006). Figure 1a represents the front view of the 3D computational domain of size 30D × 24D × 4D in the Cartesian coordinate system. The square cylinder is positioned at an upstream distance of 10D from the inlet and a downstream distance of 20D from the outlet boundary in the X direction and an equal distance of 12D from the top and bottom walls in the Y direction. The length of the cylinder is oriented along the Z direction as shown in Fig. 1(b).

Table 1 Results from the mesh convergence study performed to determine mesh independence of the solution for the flow past an unmodified square cylinder at Re = 1 × $10^6$

| Mesh No: | Mesh 1 | Mesh 2 | Mesh 3 | Mesh 4 |
|---|---|---|---|---|
| Mesh Count | 633270 | 2574667 | 4946393 | 5291319 |
| $C_d^{mean}$ | 2.76 | 2.25 | 2.33 | 2.33 |
| $C_d^{rms}$ | 0.21 | 0.27 | 0.23 | 0.20 |
| $C_l^{rms}$ | 1.40 | 1.15 | 1.30 | 1.33 |

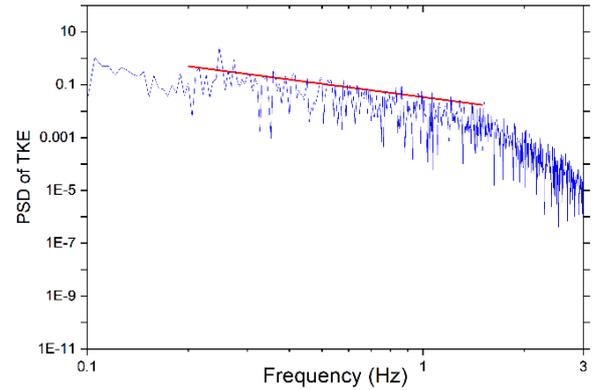

Fig. 3 Plot of Power Spectral Density of Turbulent Kinetic Energy $\frac{1}{2}\overline{u'u'}$ for point (0.75D, 0). The red line represents a line with -5/3 slope, as given by Kolmogorov's -5/3 power law

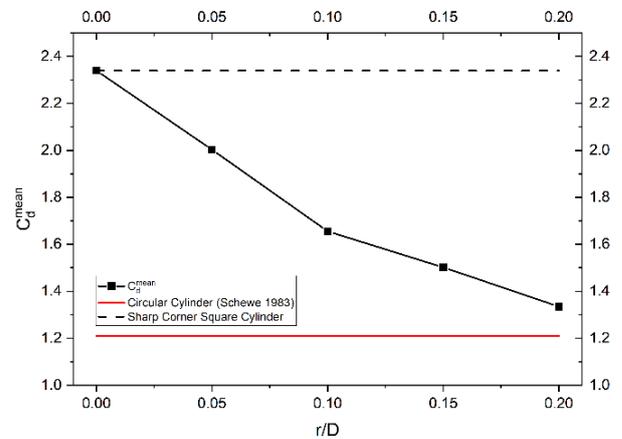

Fig. 4 Variation of $C_d^{mean}$ against normalised corner radius r/D

To better characterise the flow, we define non-dimensional numbers like the Reynolds number Re=$\rho U_0 D/\mu$, normalised corner radius r/D, and blockage ratio $B_f=D/L_y$, where D is the diameter of the square-cylinder, $L_y$=24D is the width of the domain, ρ is the density, and the dynamic viscosity of the fluid is denoted by μ. For the present case, the blockage ratio is 4.16%. A lower blockage ratio ensures that the solution is not affected by the domain boundaries. Fig. 2 shows the geometrical comparison between square cylinders with sharp and rounded corners. We simulate 4 cases with different corner modifications. The normalised corner radii considered are r/D = 0.05, 0.10, 0.15, and 0.20 in addition to the sharp-corner case.

Fig. 1 also presents the boundary conditions enforced on the domain boundaries. At the inlet boundary $\Gamma^f_{in}$, the flow enters with an uniform velocity $U_0$ along the X axis. The outlet $\Gamma^f_{out}$ has a zero-gradient velocity and zero static pressure boundary imposed on it. At $\Gamma^f_{top}$, $\Gamma^f_{bot}$, $\Gamma^f_{right}$, and $\Gamma^f_{left}$ a symmetry boundary condition has been imposed as shown in Fig. 1 due to the span-wise symmetry of the cylinder. Furthermore, the surface of the square cylinder has non-slipping walls. The values for $k_0$ and $\omega_0$ used in the inlet and outlet boundaries have been calculated based on the recommendation of (Spalart and Rumsey 2007). The corresponding turbulent intensity at inlet comes out to be 0.1%.

A mesh convergence study has been carried out to ensure that the solution is independent of any further grid refinement. Based on the results from the study, presented in Table 1, an unstructured polyhedral mesh of about 5.3 million elements has been selected. Inflation layers with a first-cell height of 0.0002D have been provided along the surface of the cylinder to satisfy the requirement of y+ ≤ 1 for the k − ω SST RANS model. To ensure that the mesh captures the characteristics of the turbulent wake adequately, the power spectra of turbulent kinetic energy is plotted against frequency for a point in the first wake region of the cylinder (0.75D, 0). In Fig. 3, the decay in energy follows a -5/3 slope, in accordance to the Kolmogorov's -5/3 power law for locally homogeneous turbulent flows. The power law is usually applicable in the inertial range and states that TKE(k) ∝ k $^{(-5/3)}$. Using the frozen turbulence hypothesis by Taylor (Taylor 1938), the decay of turbulent energy can be plotted against the frequency domain (Maryami *et al*. 2019, Gurugubelli and Jaiman 2019). Through Fig. 3, we can prove that the mesh resolution is sufficient.

In the present work, all the simulations have been carried out on FLUENT, the commercial CFD solver by ANSYS. Section 2.1 describes the finite volume scheme that discretizes the governing equations. Furthermore, the QUICK scheme (Hayase *et al*. 1992) has been used for the spatial discretization, and a bounded second order time discretization with a timestep of 0.1 D/$U_0$ has been adopted for the time marching. A Strouhal number calculation indicates that the timestep used in the current work resolves each wavelength by about 70 points which is more than sufficient to ensure accuracy in time. The results obtained from these simulations are compared with the multiple

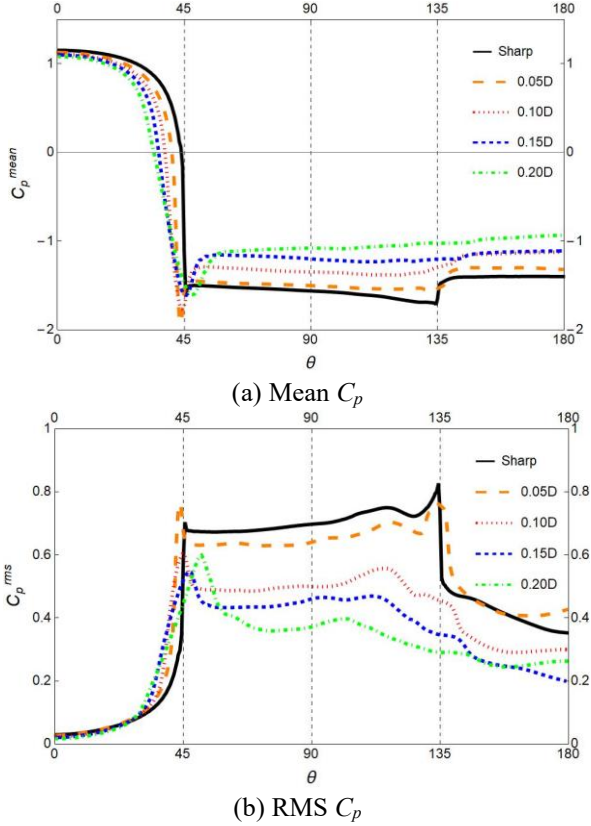

(a) Mean $C_p$

(b) RMS $C_p$

Fig. 5 Variation of mean and RMS pressure coefficient values along the surface of the cylinder for the sharp corner and rounded corner modified cases. Here θ represents the angle along the cylinder surface measured starting from the upstream stagnation point

reference works in Table 2. We observe that the drag value decreases when there is an increase in the value of turbulence intensity (Saathoff and Melbourne 1999, McLean and Gartshore 1992). In addition to turbulence intensity, a difference in the blockage ratio values also produces variations in the drag value observed for comparable Reynolds numbers (Lee 1975, Tamura and Miyagi 1999).

## 3. Results

A series of Detached-Eddy Simulations were systematically performed for a Reynolds number of Re=1 × $10^5$, considering four different normalised corner radii of r/D=0.05, 0.10, 0.15, and 0.20 to determine the effects that corner modifications have on fluid dynamic parameters such the lift coefficient, drag coefficient, Strouhal number, and the pressure coefficient. A dimensionless time-step $\delta tU_0/D$=0.10 was considered for the numerical simulations and the cases were averaged over 1500 time-step.

### 3.1 Aerodynamic drag and pressure

Figs. 4 and 5 summarise the trend observed in the drag and pressure coefficients due to the introduction of corner modifications on a square cylinder. In these figures, r/D=0 corresponds to a sharp-corner square cylinder. From Figure 4, we study the time-averaged drag coefficient ($C_d^{mean}$) for different cylinder configurations and compare them with the result obtained experimentally for a fully circular cylinder at the same Reynolds number as our simulations (Schewe 1983). $C_d^{mean}$ decreases monotonically with an increase in r/D, with a steeper drop observed between r/D = 0 and 0.10 as compared to the latter half. We observe a decrease of almost 43% in $C_d^{mean}$ between the sharp-corner case and the rounded corner cylinder with r/D = 0.20. For the corner modified cylinders, the decrease in $C_d^{mean}$ is primarily due to a decrease in the size of the stagnation pressure region in the front of the cylinder and a smaller suction pressure in the wake which can be observed in the figures 5 and 6, detailed discussions on which have been presented later in this section. Comparing the above results with those obtained for a fully circular cylinder, we observe that the $C_d^{mean}$ for a sharp-corner square cylinder is 93.32% higher than that for the fully circular case. However, the $C_d^{mean}$ for the corner modified cylinder with r/D = 0.20 is only 10.28% higher than its fully circular counterpart, thus indicating that the r/D = 0.20 configuration performs the best among the simulated cases. The variation of various force and pressure coefficients against increasing normalised corner radius indicates that small changes in the corner geometry can have substantial effects on the aerodynamic response of a square cylinder.

To explain the observations obtained in Fig. 4, we study the pressure distribution on the cylinder surface. The mean and fluctuating pressure coefficients are plotted against an angle θ with respect to the X axis, measured from the centre of the cylinder and presented in Fig. 5(a) and 5(b). In Fig. 5(a), we can notice that for a sharp corner square cylinder, $C_p^{mean}$ has two valleys at the windward and leeward corners. The valleys represent local minima in the $C_p^{mean}$ values. For the sharp corner case, the first valley is observed exactly at 45°. Comparing the pressure distribution for different r/D configurations along the windward corner, we observe that the case with r/D=0.05 has the largest dip in $C_p^{mean}$ value. This minimum occurs at an angle less than 45° and then shifts to the right along the θ axis with further modification to the corner geometry. For cases with r/D=0.15 and 0.20, the valley is observed at an angle greater than 45°. In addition to this, we also observe that the width of the valleys at the frontal corner increase with an increase in the corner radii.

Along the leeward corner, a valley is only observed for the sharp cylinder and the r/D=0.05 case. The minimum is the largest for the sharp corner configuration, and is less pronounced for the r/D=0.05 case. For the other corner modified cases, no valley is observed, thus indicating the absence of a prominent suction region on the back face of the cylinder. We note that the mean pressure at the leeward face (θ > 135°) rises as the r/D value is increased. As a result, the pressure difference across the cylinder faces decreases for higher values of r/D. Thus, a reduction in the mean drag coefficient is observed as the corner radius is increased, as observed in Fig. 4.

We also study the impact that corner modifications have the unsteady aerodynamic loads experienced by the cylinder. Fig. 5(b) illustrates $C_p^{rms}$, which is a measure of the fluctuations in the distribution of pressure along the surface of the cylinder. In Fig. 5(b), two peaks in the $C_p^{rms}$ values are observed along the surface of the cylinder that represent the pressure fluctuation amplitudes at the windward and leeward corners. For the square cylinder case with no modifications, $C_p^{rms}$ has a larger peak at the leeward corner. It then gradually decreases along the rear face of the cylinder. When a corner modification with r/D=0.05 is made, the pressure fluctuation at the windward corner becomes larger than that at the leeward corner. Furthermore, we observe a larger value of fluctuation amplitude on the back end of the cylinder for the r/D = 0.05 case as compared to others. Apart from these, $C_p^{rms}$ along the side faces for the corner modified cases are much lower than that observed for the sharp corner case. For the corner modified configurations with r/D = 0.10, 0.15, & 0.20, no peaks were observed at the leeward corner and they depart significantly from the trend observed for the sharp corner case. These observations indicate that drag reduction and suppression of unsteady aerodynamic forces are simultaneously realised by increasing the rounding radius of the corners of a square cylinder.

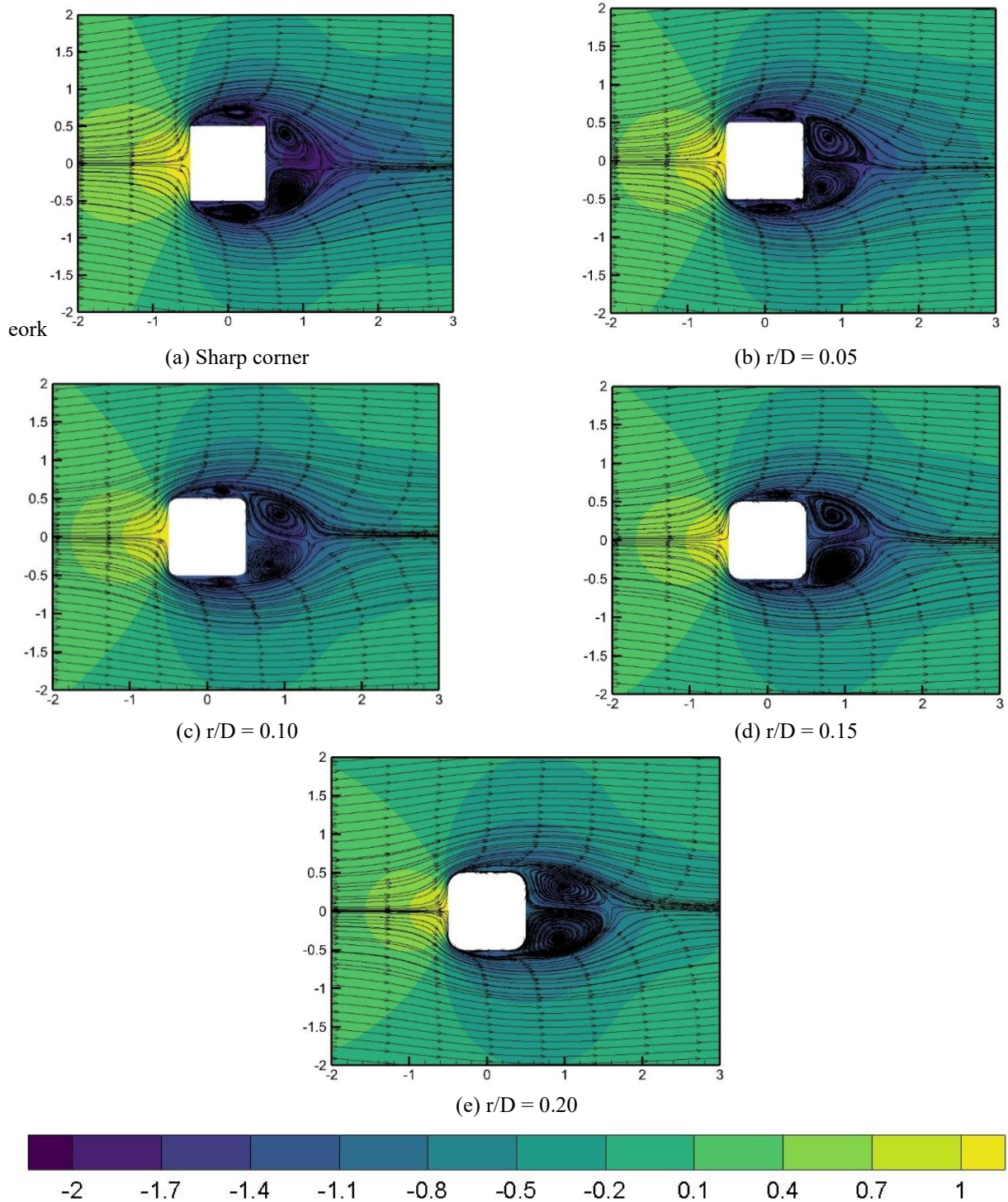

Fig. 6 Velocity streamlines superimposed on mean pressure coefficient contours for (a) the sharp case and (b)-(e) corner modified cases

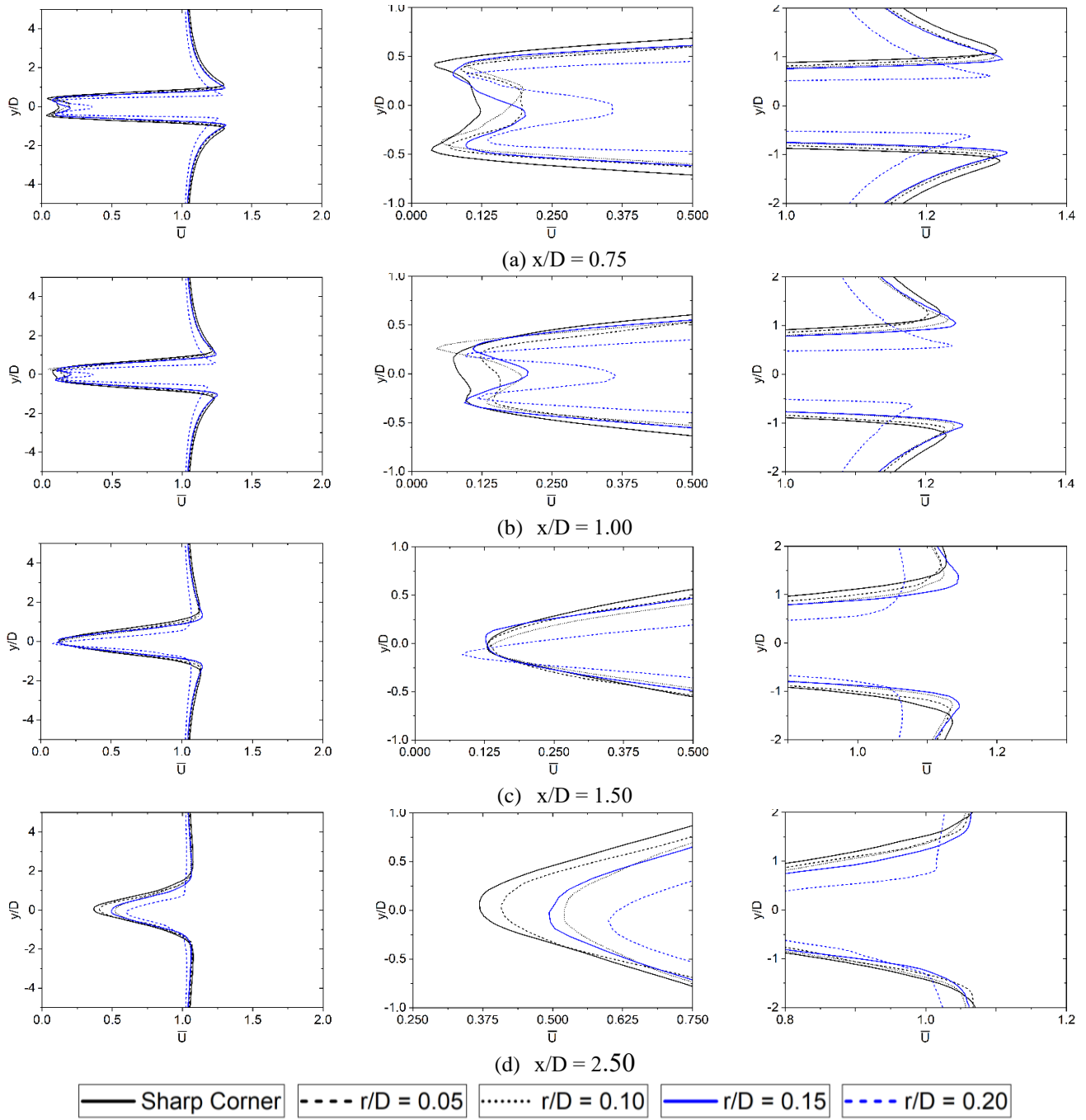

Fig. 7 Profiles of mean velocity magnitude plotted perpendicular to the Incoming flow at different points in the wake of the square cylinders

The significant effect of the normalised corner radius indicates changes in the flow patterns between the unmodified and the modified cases. We study statistical flow fields in order to explain the characteristics of local pressure distribution observed in Fig. 5. Fig. 6 represents the span and time-averaged velocity streamline plots superimposed on the mean pressure coefficient contours for the five simulated cases. A major difference in the flow topology for the five cases is caused by a shift in the flow separation point for the corner modified cases. When the rounding radius of the corners is increased, the streamlines near the corners become more curved. As a result of this increased curvature, the streamlines bunch up together near the windward and leeward corners. This explains the valleys observed in Fig. 5(a) and the increase in the width of the valleys for higher values of r/D. For the sharp corner square cylinder, the flow separates exactly at 45°. When a corner modification of r/D=0.05 is introduced, the flow separates at an angle less than 45°. However, it increases for a subsequent increase in r/D, thus representing a corresponding delay in flow separation. We also observe a valley along the leeward corner, indicating flow separation caused by the recirculating flow in the cylinder wake. The flow separation creates a small bubble which merges into the bigger recirculation bubble created due to the initial separation at the windward corner. The creation of these smaller bubbles is indicated by a decrease in the valley size and their subsequent absence in Fig. 5(a). When the normalised corner radius is equal to 0.20, the flow creeps along the leeward corner and no bubble is formed. It can also be noted that with subsequent modifications to the corner geometry, the width of the recirculating flow on the top and bottom walls decreases. It is accompanied by a reduction in the total size of the recirculation region in the cross-flow direction and an elongation along streamwise direction. The narrowing of the recirculating region is responsible for the decrease in the mean pressure difference along the upper and lower faces of the cylinder. Upon focusing on the mean pressure distribution on the cylinder surface, we observe that the magnitude and size of the stagnation pressure envelope in front of the cylinder and the negative pressure envelope behind the cylinder decrease with a corresponding increase in the r/D value, thereby corroborating the observations made earlier.

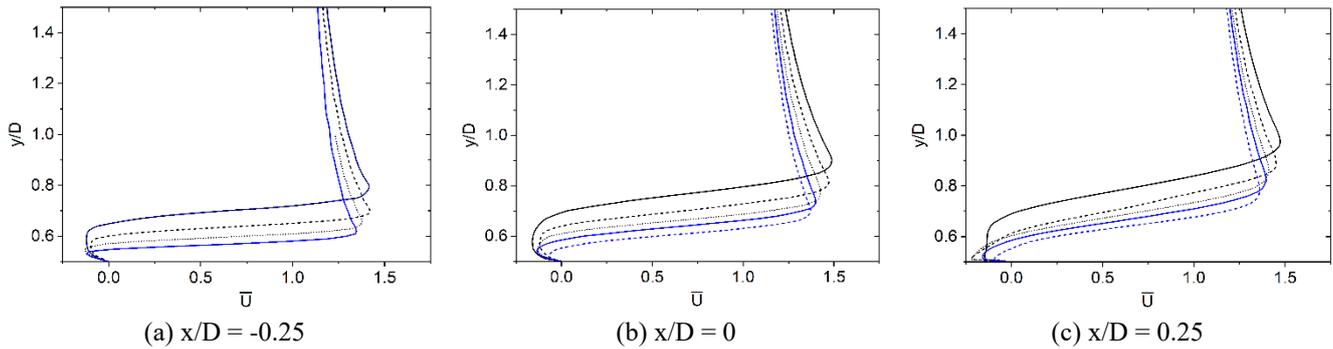

(a) x/D = -0.25  (b) x/D = 0  (c) x/D = 0.25

Fig. 8 Profiles of mean velocity magnitude plotted at three points along the top face of the cylinder. The points are situated inside the shear layer, formed due to the separation of the fluid at the frontal edge of the cylinder

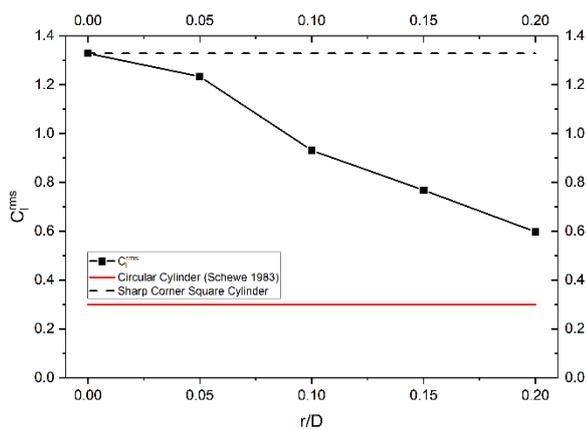

Fig. 9 Variation of fluctuating lift coefficient $C_l^{rms}$ against normalised corner radius r/D

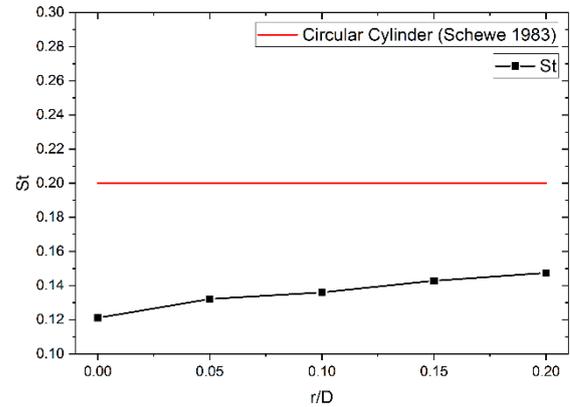

Fig. 10 Variation of Strouhal Number St against normalised corner radius r/D

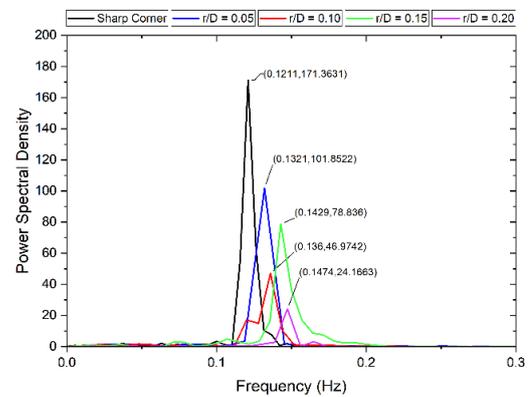

Fig. 11 Power spectral density vs Frequency plot of $C_l^{rms}$ for various square cylinder configurations

To further explain our observations from Fig. 6, we investigate the flow field behind the cylinder. Fig. 7 summarises the profiles of mean velocity obtained in the near wake (x/D=0.75 and 1.00) and the far-wake region (x/D=1.50 and 2.50). Here, x/D is the non-dimensional distance in the streamwise direction measured from the centre of the cylinder. The outward bulge in the profiles denote regions of velocity deficit where the fluid slows down in the presence of a recirculation region. In the near-wake region, we find noticeable cusps at y/D=0 and the size of these cusps decrease as we go further downstream, before completely disappearing for x/D=1.50 and 2.50. We observe that these cusps disappear quickly for the sharp corner case than in comparison to the corner modified cases, thus indicating a faster recovery of velocity deficit by the sharp corner cylinder in the near-wake region. However, the corner modified cases can better recover the velocity deficit in the far-wake, despite having a longer recirculation region, as observed in Fig. 7. This is primarily due to higher value of fluid momentum observed in the recirculation region of the corner modified cases due to the narrowing of the recirculation region in the Y direction.

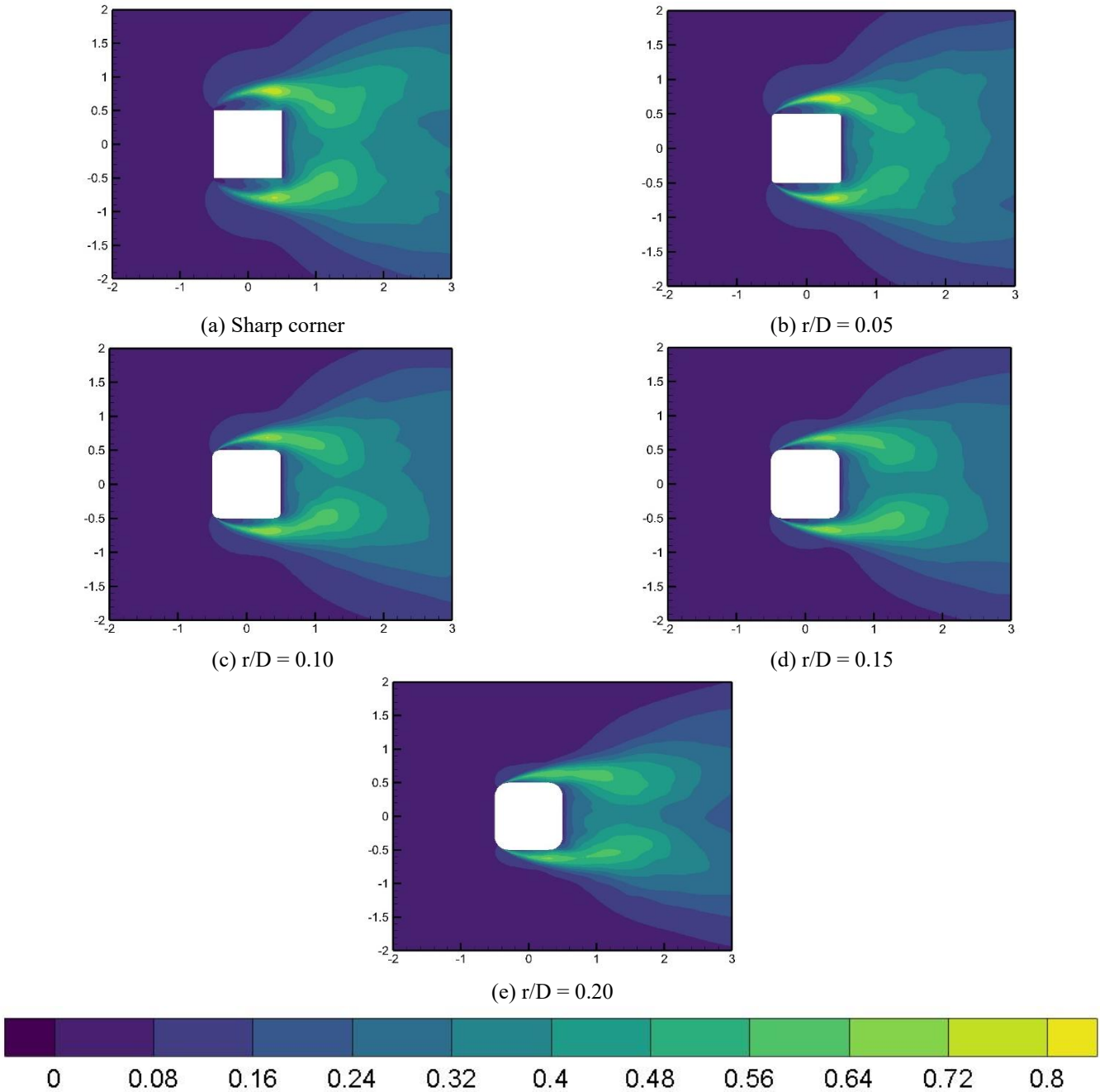

Fig. 12 Span-averaged fluctuating stream-wise velocity contours for (a) the sharp case and (b)-(e) the corner modified cases

Fig. 8 presents the streamwise mean velocity profile as we go away from the top surface for x/D=-0.25, 0, and 0.25.

From the figure, it can be observed that as the nondimensional corner radius is increased from 0 to 0.2, the recirculation bubble above the square cylinder becomes thinner. This is caused by a shift in the point of flow separation further downstream with respect to the cylinder center. This corroborates with the observation presented for the mean velocity streamlines in Fig. 6. This helps us understand the effect of corner radii on the shear layer region above the top surface of the square cylinder. The change in shear layer phenomena can be attributed as one of the main reasons behind the narrowing of the recirculation region in the cross-stream direction behind the square cylinder that is presented in Fig. 5. The cross-stream distance that corresponds to the maximum streamwise velocity also decreases with an increase in the corner radii. Thereby, it can be inferred that the effect of Kelvin-Helmholtz (KH) instability decreases. This weakens the KH vortices in the shear layers above and below the square cylinder (Lander *et al.* 2016, Cruz *et al.* 2022).

### *3.2 Aerodynamic lift and strouhal number*

In this section, we focus on the characteristics of the wake by examining the variations produced in the Strouhal number and the lift coefficients by the corner modifications. We start by considering the root mean squared (RMS) values of instantaneous lift coefficient ($C_l^{rms}$) presented in Fig. 9. $C_l^{rms}$ is a measure of the fluid force exerted on the surface of the cylinder that causes oscillations in the cross-flow direction. $C_l^{rms}$ is highest for the sharp corner case and decreases monotonically as the r/D value is increased for the corner modified cases. The high value of $C_l^{rms}$ for the sharp corner square cylinder can be attributed to the size of the wake in the Y direction. There is a nearly 55% decrease in lift coefficient between the sharp corner and the r/D=0.20 case, caused by a narrowing of the wake region. In particular, a sharp drop is observed between the r/D = 0.05 and r/D = 0.10 cases. Comparing the square cylinder cases to the fully circular cylinder (Schewe 1983), we note that $C_l^{rms}$ for the circular cylinder is 77.4% lower than that for the sharp corner case, and approximately 49.82% lower than that of the corner modified case with r/D = 0.20.

The time histories of the lift coefficient were subjected to a spectral analysis to understand the frequency of wind loads experienced by the cylinder. A Fast Fourier Transform (FFT) was performed to obtain the Strouhal number, denoted by St=f D/$U_0$. Here, f represents the frequency of the shed vortices, the characteristic dimension of the flow is denoted by D, and $U_0$ is the freestream velocity. The variation of Strouhal number against an increase in the normalised corner radius is presented in Fig. 10. A comparison between the simulated cases and a simple circular cylinder is also provided by Fig. 10 (Schewe 1983). We observe that St increases monotonically for the square cylinder cases, with the sharp-corner square cylinder having the lowest value. Compared to the sharp corner case, the St value for r/D=0.20 configuration increased by 22.4%. This increase can be attributed to the delay in flow separation at the rounded corners, which give rise to narrower, but longer recirculation regions in the cylinder wake as observed in Fig. 6. This leads to the vortices being shed at a higher frequency. Therefore, we can infer that the size of the wake region along the cross-flow direction directly influences the values are still low in comparison to the value observed for a simple circular cylinder (Schewe 1983). The Strouhal number for the r/D = 0.20 case is 25.8% lesser than the results obtained for its fully circular counterpart.

Fig. 11 displays the energy spectra for the RMS values of the lift coefficient for the various square cylinder cases. The peaks represent the energy of the shed vortices in the wake. We note that the magnitude of the peaks decreases for higher degrees of modification. Instead of a monotonic decrease, we observe a slight increase in the shedding strength upon increasing the normalised corner radius from 0.10 to 0.15. The case with r/D=0.20 has the weakest shedding, approximately 86% weaker than the sharp corner case. In Fig. 11, the X axis represents the dominant frequency with which vortices are shed in the wake. As the corner rounding radius is increased, the peaks shift further right along the X axis, thereby corroborating the observations obtained from Fig. 10.

The turbulent statistics of the wake are further studied using the span-averaged fluctuating stream-wise velocity contours presented in Fig. 12. We observe that for all simulated cases, the velocity fluctuation in the wake region is lower than that along the side walls. We also observe that the magnitude of peak velocity fluctuations decreases with an increase in the r/D value. Furthermore, as we move away from the body, the effect of streamwise velocity diminishes. It is visible in the form of a decreasing region of influence with progressive corner modifications. The peaks are generated because the shear layers separate from the frontal edge of the cylinder and get deflected at large angles. The magnitude of the peaks decrease as the wake becomes narrower in the cross-flow direction. A reduced fluctuating stream-wise velocity and a reduced region of influence lead to a decrease in the RMS values of the force coefficients, as has been observed in the case of $C_l^{rms}$ in Fig. 9.

## 4. Conclusions

This study presents an extensive analysis of the effects of rounded corner modifications 0 ≤ r/D ≤ 0.20 on a square cylinder's aerodynamic response at a practical Reynolds number of Re=1 × $10^5$. 3D simulations have been carried out using the IDDES formulation and validation of results against published LES simulations of the flow around sharp-corner square cylinders are provided. The conclusions can be summarised as follows:

• Increasing the corner radius leads to a steady decrease in separation angle. It is accompanied by a narrowing of the recirculation region near the leading edge and a corresponding increase in its length along the streamwise direction.

• Due to geometric modifications to the corners, the pressure across the cylinder faces in the streamwise direction reduces, causing a drop in the mean drag.

• Considering the aerodynamic force coefficients, a substantial decrease in their mean and RMS values is observed when corner modifications are introduced. Specifically, a 43.2% decrease in $C_d^{mean}$, a 62.1% decrease in $C_d^{rms}$, and a 55.5% decrease in $C_l^{rms}$ between the sharp case and the r/D = 0.20 case is observed.

• The elongation of the recirculation region resulted in a higher frequency of vortex shedding downstream of the cylinder. It also led to a faster recovery of the velocity deficit farther in the far-wake of the cylinder.

The technical implications of the observations above are important in studies that aim to decrease wind loading on bluff bodies like short buildings, which are subject to high Reynolds number flows. This numerical simulation provides evidence of significantly reduced aerodynamic force coefficients upon increasing the corner modification

## Acknowledgments

The authors sincerely acknowledge the financial support received from LIGO India to carry out the following study.

AD